\documentclass[aps,floatfix,prd,superscriptaddress,reprint,showpacs,10pt,preprintnumbers,longbibliography]{revtex4-2}
\usepackage[utf8]{inputenc}
\usepackage[pdftex]{graphicx}
\usepackage{float}
\usepackage{amssymb}
\usepackage{amsmath}  
\usepackage{dsfont}
\usepackage{array}
\usepackage{bm}
\usepackage{mathrsfs}
\usepackage{pifont}
\usepackage{multirow}
\usepackage{upgreek}
\usepackage{xcolor}
\usepackage[pdftex,
            pdftitle={Phase structure of the CP(1) model in the presence of a topological term},
            pdfauthor={Katsumasa Nakayama, Lena Funcke, Karl Jansen, Ying-Jer Kao, Stefan Kuehn},
            bookmarks,
            colorlinks,
            linkcolor=myblue,
            citecolor=mymagenta,
            menucolor=black,
            urlcolor=myblue,
            plainpages=false,
            pdfpagelabels,
            hypertexnames=false]{hyperref}
\usepackage{verbatim}
\usepackage{slashed}
\usepackage{cleveref}

\definecolor{mymagenta}{RGB}{200, 0, 100}
\definecolor{myblue}{RGB}{45, 48, 146}
\definecolor{mygreen}{RGB}{0, 126, 0}
\definecolor{myorange}{RGB}{255, 136, 19}

\newcommand{\kmax}{\ensuremath{k_\mathrm{max}}}

\def\be#1{\begin{equation}#1\end{equation}} 

\def\beqnn#1{\begin{eqnarray}#1\end{eqnarray}}

\begin{document}
\title{Phase structure of the CP(1) model in the presence of a topological $\theta$-term}
\date{\today}
\begin{abstract}
We numerically study the phase structure of the CP(1) model in the presence of a topological $\theta$-term, a regime afflicted by the sign problem for conventional lattice Monte Carlo simulations. Using a bond-weighted tensor renormalization group method, we compute the free energy for inverse couplings ranging from $0\leq \beta \leq 1.1$ and find a CP-violating, first-order phase transition at $\theta=\pi$. 
In contrast to previous findings, our numerical results provide no evidence for a critical coupling $\beta_c<1.1$ above which a second-order phase transition emerges at $\theta=\pi$ and/or the first-order transition line bifurcates at $\theta\neq\pi$. If such a critical coupling exists, as suggested by Haldane's conjecture, our study indicates that is larger than $\beta_c>1.1$. 
\end{abstract}

\author{Katsumasa Nakayama}
\affiliation{NIC, DESY, Platanenallee 6, D-15738 Zeuthen, Germany}

\author{Lena Funcke}
\affiliation{Perimeter Institute for Theoretical Physics, 31 Caroline Street North, Waterloo, Ontario, N2L 2Y5, Canada}

\author{Karl Jansen}
\affiliation{NIC, DESY, Platanenallee 6, D-15738 Zeuthen, Germany}

\author{Ying-Jer Kao}
\affiliation{Department of Physics and Center for Theoretical Physics, National Taiwan University, Taipei 10607, Taiwan}
\affiliation{Physics Division, National Center for Theoretical Science, National Taiwan University, Taipei 10607, Taiwan}

\author{Stefan K\"uhn}
\affiliation{Computation-Based Science and Technology Research Center, The Cyprus Institute, 20 Kavafi Street, 2121 Nicosia, Cyprus}

\maketitle
\section{Introduction}
The CP($N-1$) models in 1+1 dimensions share many properties with QCD in 3+1 dimensions, among them
confinement, asymptotic freedom, instantons, a $1/N$ expansion, a topological charge, and a $\theta$-term. Thus, they serve as benchmark models for developing and testing both new numerical techniques and new proposed solutions to open questions of QCD. 

Both the CP($N-1$) models and QCD contain many nonperturbative phenomena, which cannot be addressed with conventional lattice techniques. One particular example is the above-mentioned $\theta$-term, which is the origin of the strong CP problem in QCD~\cite{Crewther:1979pi,Abel:2020gbr}. The $\theta$-term gives rise to an imaginary contribution to the action in the Euclidean formulation of lattice gauge theories. This complex action problem, also called the sign problem~\cite{Troyer2005}, prevents the successful application of Markov chain Monte Carlo (MCMC) methods for large values of the topological vacuum angle $\theta$. To evade this problem, new numerical techniques have to be developed. 

One particularly promising approach are tensor network (TN) techniques, which have been successfully employed to simulate lattice gauge theories in 1+1 dimensions~\cite{Banuls2013,Banuls2013a,Buyens2013, Kuehn2014,Buyens2014,Buyens2015a,Buyens2015, Banuls2016a,Banuls2016b,Zapp2017,Buyens2017, Banuls2018a,Saito2014,Banuls2015,Saito2015, Buyens2016,Banuls2016,Rico2013,Pichler2015,Ercolessi2018,Magnifico2019,Magnifico:2019ulp,Kuehn2015,Banuls2017,Sala2018,Sala2018a,Silvi2016,Silvi2019,Shimizu2014,Shimizu2014a,Kawauchi:2016dcg,Kawauchi:2017dnj, Byrnes2002,Buyens2017,Funcke:2019zna,Bruckmann:2018usp} and have recently been applied to theories in 2+1 \cite{Tagliacozzo:2012vg,Tagliacozzo:2014bta,Zohar:2015eda,Kuramashi2018,Felser:2019xyv,Emonts:2020drm} and 3+1~\cite{Magnifico:2020bqt} dimensions. In particular, the $\theta$-dependence of lattice gauge theories has already been successfully studied using various TN methods, i.e., the density matrix renormalization group method~\cite{Byrnes2002}, matrix product states~\cite{Buyens2017,Funcke:2019zna}, and the tensor renormalization group (TRG) approach~\cite{Kawauchi:2016dcg,Kawauchi:2017dnj,Shimizu2014a}. Specifically, TRG has been used to study the $\theta$-dependence of the CP(1) model~\cite{Kawauchi:2016dcg,Kawauchi:2017dnj} and the Schwinger model~\cite{Shimizu2014a}.

\begin{figure}[t!]
  \centering
  \includegraphics[width=0.95\columnwidth]{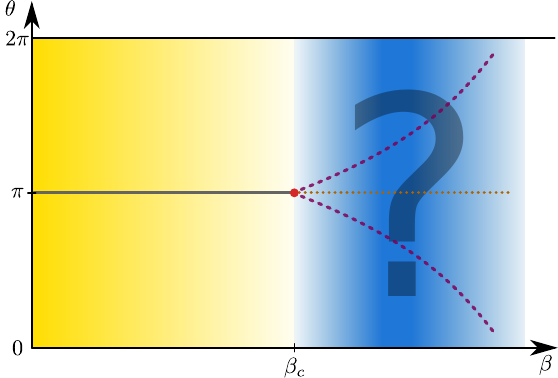}  
  \caption{Sketch of the phase diagram of the lattice CP(1) model, where $\beta=1/g^2$ is the inverse coupling and $\theta$ is the free parameter of the topological $\theta$-term. The weak-coupling limit $\beta\to\infty$ corresponds to the continuum limit of the lattice model. For small values of $\beta$, the model undergoes a first-order phase transition at $\theta=\pi$ (solid gray line), up to a certain critical coupling $\beta_c$ (yellow shaded region). For $\beta>\beta_c$ (blue region), different predictions exist and the phase structure is not entirely clear. There are hints toward a second-order phase transition at $\theta=\pi$ (dotted brown line), in agreement with Haldane's conjecture, and toward a bifurcation of the critical first-order line, such that the model undergoes two first-order transitions in $\theta\neq\pi$ (dashed dark purple lines).}  
  \label{fig:phase_diagram}
\end{figure}

The phase structure of different CP($N-1$) models has been intensively investigated with different methods in the past~\cite{Schierholz:1994pb, Plefka:1996tz,Plefka:1996ks,Imachi:1999xt, Azcoiti:2007cg,Kawauchi:2016dcg,Kawauchi:2017dnj}. In particular, the phase diagram of the CP(1) model with a $\theta$-term has been studied with a strong coupling analysis~\cite{Plefka:1996ks}, Monte Carlo simulations~\cite{Azcoiti:2007cg}, and TRG studies~\cite{Kawauchi:2016dcg,Kawauchi:2017dnj}. Here, it was shown that a first-order phase transition occurs at small values of $\beta$ and $\theta=\pi$, where the CP symmetry spontaneously breaks (see Fig.~\ref{fig:phase_diagram}). For larger values of $\beta$, corresponding to the weak-coupling regime, the numerical studies in Refs.~\cite{Azcoiti:2007cg,Kawauchi:2016dcg,Kawauchi:2017dnj} indicated that the phase transition at $\theta=\pi$ becomes a second-order transition. Since the CP(1) model is equivalent to the O(3) model~\cite{Banerjee:1994zzb}, these results are in agreement with Haldane's conjecture~\cite{Haldane:1982rj,Haldane:1983ru} that the O(3) model at weak coupling becomes gapless at $\theta=\pi$, corresponding to a second-order phase transition. At the same time, it has also been suggested that the first-order transition line bifurcates at $\theta\neq \pi$ for large values $\beta$, thus exhibiting a non-trivial dependence on the coupling~\cite{Plefka:1996ks,Azcoiti:2007cg,Kawauchi:2016dcg,Kawauchi:2017dnj}. However, the exact value of the critical coupling, $\beta_c$, remains unknown for both scenarios. While a strong coupling analysis suggests a bifurcation at $\beta_c=0.56$~\cite{Plefka:1996ks}, Monte Carlo simulations yield a second-order transition beyond $\beta_c = 0.5$~\cite{Azcoiti:2007cg}, and the TRG analysis finds that the second-order phase transition occurs around $\beta_c\approx0.3-0.4$~\cite{Kawauchi:2016dcg,Kawauchi:2017dnj}.

In the present work, we reanalyze the phase diagram of the lattice CP(1) model in the presence of a $\theta$-term using high-precision TRG simulations. We provide a detailed analysis of the different errors contributing to the current and previous numerical studies. Interestingly, we find no indication of a bifurcation point or the onset of a second-order phase transition up to $\beta=1.1$, which is much larger than suggested by the previous studies. If the bifurcation point really exists, as suggested by Haldane's conjecture, our results indicate it occurs at $\beta_c>1.1$. 

The paper is organized as follows. In Sec.~\ref{sec:model}, we will describe the CP(1) model and the numerical TRG methods. In Sec.~\ref{sec:results}, we will present our numerical results. In Sec.~\ref{sec:conclusion}, we will summarize and discuss our results.

\section{Model and methods\label{sec:model}}
\begin{figure}[t!]
  \includegraphics[width=0.75\columnwidth]{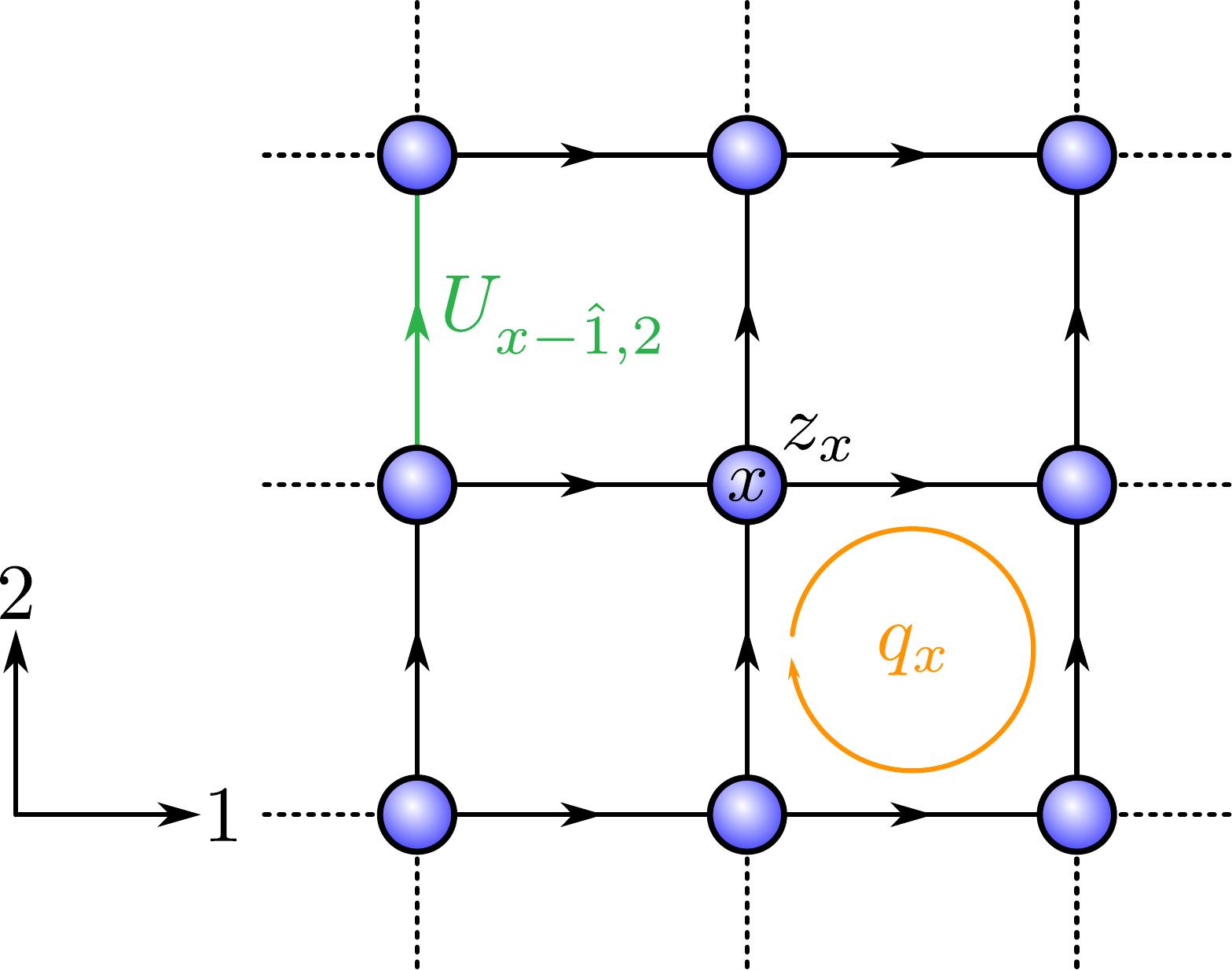}
  \caption{Illustration of the 1+1D square lattice showing the vertices (blue spheres) and the directed links. The complex scalar fields $z_x$ (illustrated in black) are sitting on the vertices $x$ and the link variables $U_{x,\mu}$ are sitting on the edges connecting sites $x$ and $x+\hat{\mu}$ as illustrated in green. The topological charge $q_x$ (orange) corresponds to a product of the sum of the gauge fields around a plaquette.}
  \label{fig:lattice}
\end{figure}

\begin{figure*}
\includegraphics[width=1.0\textwidth]{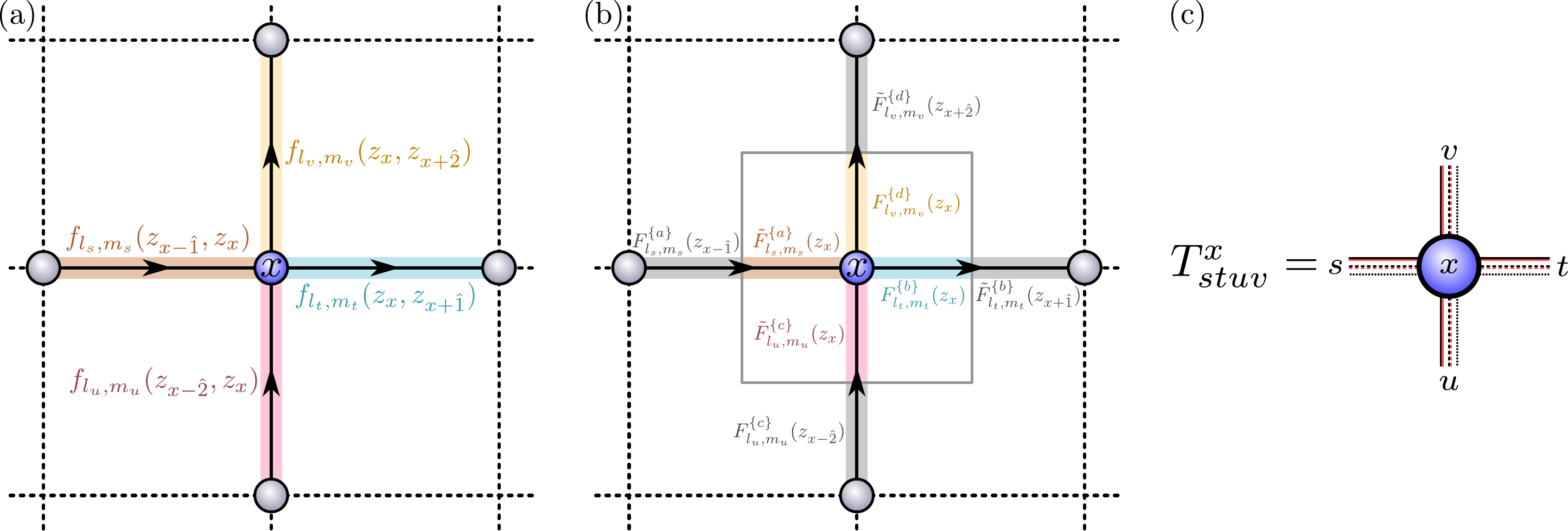}
  \caption{(a) Illustration of the function $f_{l,m}(z_x,z_{x+\hat{\mu}})$ (indicated by the shaded region behind the gauge links) connecting the complex scalar field at site $x$ (blue sphere) to its neighboring lattice sites (gray spheres). Note that each site has two ingoing and two outgoing links, hence there are four pairs of integer indices $(l_i,m_i)$, $i=s,t,u,v$ at each site. (b) Illustration of the situation after using Eq.~\eqref{eq:f_decomposition} to break the function $f_{l,m}(z_x,z_{x+\hat{\mu}})$ into a contraction of two tensors between neighboring sites. The colored part belongs to site $x$, the gray parts belong to the neighboring sites. The gray square illustrates the pieces which form the tensor $W^x$ in Eq.~\eqref{eq:Dx} upon integrating over the scalar field. Panel (c) shows the final rank-4 tensor $T^x_{stuv}$ sitting at vertex $x$. The bonds of the tensor are represented as legs sticking out of the vertex. Each bond corresponds to a multi-index corresponding to a pair of indices $(l_i,m_i)$ resulting from the character expansion (solid black and red lines), two sets of indices $\{a\}$ from the decomposition of the function $f_{l_i,m_i}$ in Eq.~\eqref{eq:decomposition} (dashed black and red lines), and an index from the expansion of topological $\theta$-term (black dotted line).}
  \label{fig:tensor}
\end{figure*}

In order to study the phase structure of the CP(1) model with the TRG~\cite{Levin:2006jai,Xie2012,Kawauchi:2016xng}, we use a Euclidean time lattice formulation. The discretized action of the model is given by
\be{
S_\theta
=
-2\beta\sum_{x,\mu}
\left[
z^*_x z_{x+\hat{\mu}}U_{x,\mu}
+
z_xz^* _{x+\mu} U^\dagger_{x,\mu}
\right]
-
i\frac{\theta}{2\pi}\sum_{x}q_x,
\label{eq:S}
}
where $\beta=1/g^2$ is the inverse coupling constant and $\theta\in[0,2\pi]$ is the free parameter of the topological $\theta$-term. The two-component complex scalar fields $z_x\in\mathds{C}^2$ reside on the sites $x=(x_1,x_2)$ of a two-dimensional lattice and are fixed to unit length, $z_x^*z_x = 1$ $\forall x$. The gauge variables $U_{x,\mu}$ reside on the links connecting the lattice sites $x$ and $x +\hat{\mu}$ along the direction $\mu$ (see Fig.~\ref{fig:lattice}). They are related to the components $A_{x,\mu}$ of an auxiliary vector field via $U_{x,\mu}=\exp(iA_{x,\mu})$. The topological charge $q_x$ is defined as 
\be{
q_x
=
A_{x,1}
-
A_{x+\hat{1}-\hat{2},2}
-
A_{x-\hat{2},1}
+
A_{x-\hat{2},2}
\mathrm{\ mod\ }
2\pi.
}
In order to explore the phase diagram of the CP(1) model using TRG, we need to obtain the tensor network representation of the partition function
\begin{align}
  \mathcal{Z}_{\theta} = \int \prod_x \mathrm{d}z_x  \prod_{y,\mu} \mathrm{d}A_{y,\mu}\ \exp(-S_{\theta}),
  \label{eq:partition_function}
\end{align}
where we have used a shorthand notation for the integration measure over the complex scalar fields,
\begin{align}
  \mathrm{d}z_x =  \mathrm{d}^2z^*_x\ \mathrm{d}^2z_x\ \delta(|z_x|-1).
  \label{eq:measurement}
\end{align}
To this end, we first derive an expansion for $\exp(-S_{\theta=0})$ which factors into two parts: a part dependent on the complex scalar fields $z_x$ and a part depending on the vector fields $A_{x,\mu}$. Inserting this expansion into Eq.~\eqref{eq:partition_function} and taking the integral over the scalar fields, we arrive at a tensor representation of the matter part. Subsequently, we derive a corresponding representation for the $\theta$-term. The final tensors for $\mathcal{Z}_\theta$ can then be obtained by multiplying both tensors for the individual parts together and integrating over the gauge fields. To obtain a tensor network representation for $\theta=0$, we use the characterlike expansion of the Boltzmann factor~\cite{Plefka:1996ks} 
\begin{align}
  \begin{split}
    e^{-S_{\theta=0}}
    =
    \prod_{x,\mu}
    Z_0(\beta)
    \sum_{l,m = 0} ^\infty
    d_{l,m}
    \exp{[
    i(m-l)A_{x,{\mu}}]
    }\\
    \times h_{l,m}(\beta)
    f_{l,m}(z_x,z_{x+\hat{\mu}}),
    \label{eq:expS}
  \end{split}
\end{align}
where $d_{l,m}$ are the expansion dimensions, $h_{l,m}$ are the expansion coefficients, $f_{l,m}$ are the expansion characters, and $Z_0(\beta)$ is a normalization factor. The non-negative integers $l$ and $m$ will become the indices of the tensor representation after integrating out the degrees of freedom $z_x$ and $A_\mu$. Note that due to the product in front of the right-hand side of Eq.~\eqref{eq:expS}, there are four sets of indices $(l,m)$ associated with each site. These correspond to the ingoing and outgoing index for every direction [see Fig.~\ref{fig:tensor}(a)].

The expansion coefficients $h_{l,m}$ and the normalization factor $Z_0(\beta)$ can be expressed through the modified Bessel functions of the first kind, $I_n(x)$, 
\begin{align}
  h_{l,m}(\beta)
  &=
  \frac{I_{1+l+m}(4\beta)}
  {I_1(4\beta)}, \label{eq:hml}\\
  Z_0(\beta)
  &=
  \frac{I_{1}(4\beta)}
{2\beta}.
\end{align}
As the explicit form of the dimensions $d_{l,m}$ and the characters $f_{l,m}$ is rather complicated, we refer to Refs.~\cite{Plefka:1996ks,Kawauchi:2016xng} for details and only quote their normalization conditions here,
\begin{align}
  \begin{split}
    \int \mathrm{d}z_x\
    f_{l,m}(z_{x'},z_x)
    f^* _{l',m'}(z_{x''},z_x)
    &=
    \frac{1}{d_{l,m}}f_{l,m}(z_{x'},z_{x''}),\\
    f_{l,m}(z_x,z_x) &= d_{l,m},
  \end{split}
	\label{eq:f_decomposition}
\end{align}
In order to construct the tensor representation, we introduce an additional decomposition of $f_{l,m}$ [see Fig.~\ref{fig:tensor}(b)],
\be{
  f_{l,m}(z_x,z_{x'})
  =
  \sum_{\{a\}}
  F_{l,m}^{\{a\}}(z_x)
  \tilde{F}_{l,m}^{\{a\}}(z_{x'}),
  \label{eq:decomposition}
}
where $\{a\} = a_1,...,a_l, a'_1,...,a'_m$, and the indices $a_i$ ($a_i'$) range over $1,2$ (see Refs.~\cite{Plefka:1996ks,Kawauchi:2016xng} for details).

With this expansion, we can divide $f_{l,m}(z_x,z_{x'})$ into a $z_x$-dependent part $F_{l,m}^{\{a\}}(z_x)$ and a $z_{x'}$-dependent part $\tilde{F}_{l,m} ^{\{a\}}(z_{x'})$, such that we can subsequently integrate out the complex scalar fields $z_x$ site by site. Note that the $\theta$-term does not depend on $z_x$, which implies that only the first term in Eq.~\eqref{eq:S} is relevant for integrating out $z_x$. Finally, we define the tensor [see Fig.~\ref{fig:tensor}(b)] by the integral 
\begin{align}
  \begin{aligned}
    W^x& _{(l_s,m_s,\{a\})(l_t,m_t,\{b\})(l_u,m_u,\{c\})(l_v,m_v,\{d\})}\\
    \equiv&\,
    \sqrt{
    d_{l_s,m_s}
    d_{l_t,m_t}
    d_{l_u,m_u}
    d_{l_v,m_v}
    }\\
    &\times\,
    \sqrt{
    h_{l_s,m_s}(\beta)
    h_{l_t,m_t}(\beta)
    h_{l_u,m_u}(\beta)
    h_{l_v,m_v}(\beta)
    }\\
    &\times\,
    \int \mathrm{d}z_x\
    \tilde{F}^{\{a\}} _{l_s,m_s} (z_x)
    {F}^{\{b\}} _{l_t,m_t} (z_x)
    \tilde{F}^{\{c\}} _{l_u,m_u}(z_x)
    {F}^{\{d\}} _{l_v,m_v}(z_x).
    \end{aligned}
  \label{eq:Dx}
\end{align}
The tensor $W_x$ represents a rank-4 tensor, whose bonds are given by the four multi indices $(l_s,m_s,\{a\})$, $(l_t,m_t,\{b\})$, $(l_u,m_u,\{c\})$, and $(l_v,m_v,\{d\})$, which allows for representing $\exp(-S_{\theta=0})$ as a tensor network up to the factor $e^{i(m-l)A_{x,\mu}}$ depending on the gauge field.

Next, we derive a tensor representation of the topological $\theta$-term starting again from its character expansion, which reads
\be{
e^{i\frac{\theta}{2\pi} q_x}
=
\sum_{n_p\in \mathbb{Z}}
e^{in_p
q_x
}
G_{n_p}(\theta).
\label{eq:expansion_topological_term}
}
In the expression above, $n_p$ is an integer, and we have defined 
\be{
G_{n_p}(\theta)
=
\frac{2\,\mathrm{sin}\frac{\theta + 2\pi n_p}{2}}{\theta + 2\pi n_p}.
\label{eq:Gntheta}
}
In order to take the integral over the gauge field $A_{x,\mu}$, we consider the following integrals of $A_{x,1}$ and $A_{x-\hat{2},2}$,
\begin{align}
\begin{split}
\int^\pi _{-\pi}
\mathrm{d}A_{x,1}
e^{i(m_t-l_t)A_{x,1}}
e^{i(t_p-v_p)A_{x,1}}
&=
\delta^{l_t-m_t} _{t_p-v_p},\\
\int^\pi _{-\pi}
\mathrm{d}A_{x-\hat{2},2}
e^{i(m_u-l_u)A_{x-\hat{2},2}}
e^{i(s_p-t_p)A_{x-\hat{2},2}}
&=
\delta^{l_u-m_u} _{s_p-t_p}.
\end{split}
\end{align}
Similarly, the integral over $A_{x,2}$ yields $\delta ^{l_t-m_t} _{u_p-v_p}$. Using $\delta_{ab}\delta_{bc}=\delta_{ac}$, we finally get the following tensor representation,
\beqnn{
T^x _{stuv}
&\equiv&
T^x _{
(l_s,m_s,\{a\},s_p)
(l_t,m_t,\{b\},t_p)
(l_u,m_u,\{c\},u_p)
(l_v,m_v,\{d\},v_p)
}\nonumber\\
&\equiv&
W^x _{
(l_s,m_s,\{a\})
(l_t,m_t,\{b\})
(l_u,m_u,\{c\})
(l_v,m_v,\{d\})
}
\nonumber\\
&\times&
\delta^{l_t-m_t} _{t_p-v_p}
\delta^{l_u-m_u} _{s_p-t_p}
G_{t_p}(\theta)
\delta_{t_pu_p}
.
}
The partition function of the system can now be represented as the contraction of the rank-4 tensors $T_{stuv}^x$ sitting at the different lattice sites. Compared to $W^x$, the multi-index for each bond of the tensor $T^x$ comprises an additional index resulting from the character expansion of the topological term [see also Fig.~\ref{fig:tensor}(c)].

Since the indices $l$, $m$, and $n_p$ in the character expansions in Eq.~\eqref{eq:S} and Eq.~\eqref{eq:expansion_topological_term} range over a countable infinite set, the bonds of the tensor $T^x$ are infinite dimensional. Hence, for numerical calculations we need to truncate the bonds to a finite dimension. To this end, we limit the number of terms that we keep in the character expansions. For the indices $l$, $m$ we define a cutoff $\kmax$ and restrict ourselves to values fulfilling $l + m\leq \kmax$. This results in a bond size of $\chi_\beta = 1 + \kmax 2^{\kmax + 1}$ for the tensor representing the partition function without the $\theta$-term. Similarly, we only keep a subset of $\chi_\theta$ terms in the expansion for the topological term in Eq.~\eqref{eq:expansion_topological_term}. Since our numerical analysis primarily focuses on the region $\theta=\pi$, we choose the terms with the largest absolute value of $G_{n_p}(\pi)$. For example, for $\chi_\theta = 2$, we take into account $n_p = 0, -1$ because the absolute value of $|G_{0,-1}(\pi)| = {2}/{\pi}$ is larger than all other values of $|G_{n_p}(\pi)|$. The size of the truncated tensor $T^x$ is thus given by $\chi_\beta \times \chi_\theta$. 

These cutoffs in the tensor $T^x$ introduce systematic errors. Since $h_{l,m}$ is given by a ratio of the modified Bessel functions (see Eq.~\eqref{eq:hml}), the cutoff at $\kmax$ truncates only an exponentially small contribution to the tensor $T_{stuv} ^x$. The truncation in the expansion for the $\theta$-term leads to an error of the order $\mathcal{O}(1/\chi_\theta)$, as shown in Eq.~\eqref{eq:Gntheta}. We will estimate these systematic errors by using different $\kmax$ and $\chi_\theta$ our numerical calculations.

In addition, the TRG algorithm introduces another systematic error. Starting from the initial, finite dimensional tensor $T^x$, the coarse-gaining step of the algorithm  would produce a tensor which has the same structure as the original one, but with a bond size corresponding to the square of the one of the original tensor. Hence, for the computation to be sustainable, one has to truncate the tensor during the coarse-graining step and limit the bond size to a maximum value $D$. We will estimate this additional systematic error by performing calculations for various values of the bond dimensions $D$ for a fixed set of parameters. In particular, $D$ should be chosen larger than the size of the initial tensor $\chi_\beta \times \chi_\theta$ to capture the relevant physics of the model.

\section{Results\label{sec:results}}
In order to explore the phase structure of the model, we compute the partition function using TRG methods, which in turn allows us to obtain the free energy density
\begin{align}
  F(\beta,\theta) = -\frac{1}{\beta V} \log\mathcal{Z}_\theta,
  \label{eq:free_energy}
\end{align}
where $V$ is the dimensionless volume of the lattice. Moreover, we can also compute the specific heat from the partition function using the relation
\begin{align}
C(\beta,\theta) = \frac{\beta ^2}{V} \frac{\partial^2}{\partial \beta^2}\log\mathcal{Z}_\theta.
  \label{eq:specific_heat}
\end{align}
In particular, studying the free energy will allow us to determine the phase structure of the model and to detect a possible phase transition as discontinuities in the derivatives of the free energy. 

As a first step, we focus on the case $\theta=0$ to explore systematic errors due to the truncation in $\kmax$ over a large range $\beta$. This allows us to identify the range of couplings in which our TRG computations yield reliable results. As a second step, we turn to the model in the presence of a topological $\theta$-term. Using again the TRG approach, we study the model in a region around $\theta=\pi$, where we carefully estimate our systematic errors due to the finite values of $(D,\kmax, \chi_\theta, V)$.

\subsection{CP(1) model without $\theta$-term}
To begin with, we benchmark the TRG method for $\theta=0$ and explore the effect of systematic errors due to the truncation of the expansion in Eq.~\eqref{eq:expS}. To this end, we calculate the free energy and specific heat of the CP(1) model using the anisotropic TRG (ATRG) method with randomized singular value decomposition (SVD)~\cite{Morita2018,PhysRevB.102.054432}. In order to avoid errors due to the randomized SVD, we set the oversampling parameter to the value of $D$, meaning that we compute $2D$ singular values at each coarse-graining step and select the $D$ dominant ones from those. This ATRG method is faster than the usual TRG algorithm, while the precision is comparable to the one of TRG for same $D$. Since we choose the oversampling parameter large enough, the systematic error from the randomized SVD is negligible.

\begin{figure}[htp!]
  \centering
  \includegraphics[width=1.0\columnwidth]{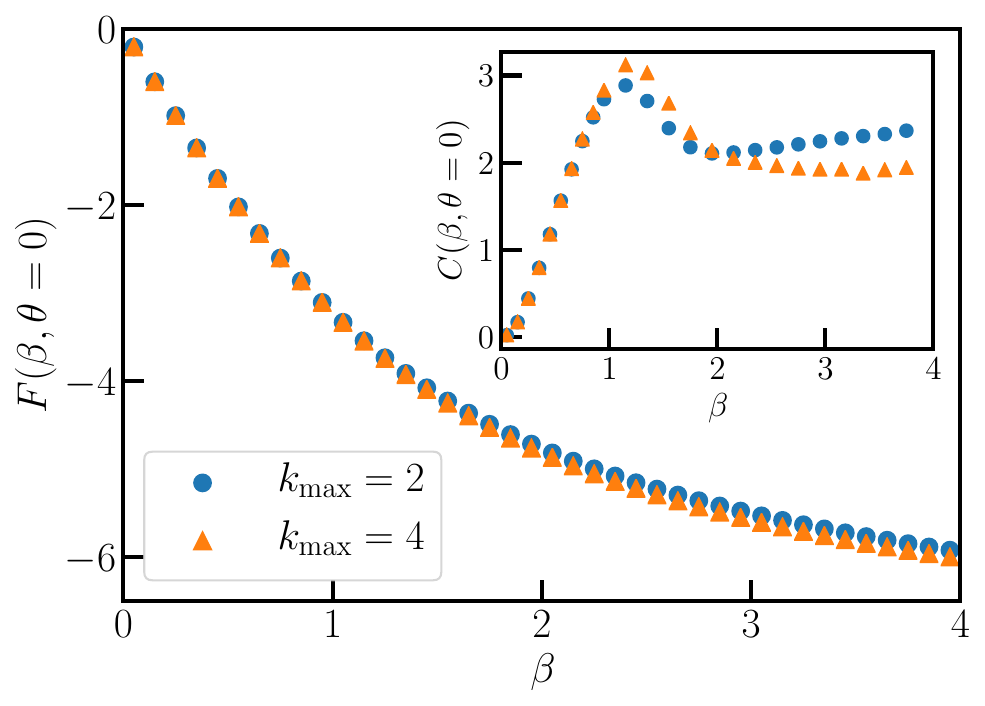}  
  \caption{ATRG results for the free energy $F(\beta,\theta=0)$ (main plot) and the specific heat $C(\beta,\theta=0)$ (inset) as a function of the inverse coupling $\beta=1/g^2$, with the volume $V=2^{40}$, bond size $D=80$, and cutoff $\kmax = 2$ (blue dots) and $4$ (orange triangles).}
  \label{fig:TRG_no_theta}
\end{figure}

Figure~\ref{fig:TRG_no_theta} shows our results for the free energy and the specific heat for $D=80$, $\kmax = 2,4$, and a volume of $V=2^{40}$.
Focusing on the free energy, we see that the systematic error due to the finite value of $\kmax$ is negligible in the range of small values of the inverse coupling. In particular, for $\beta\leq 1.5$ there is essentially no difference between results with $\kmax=2$ and $4$. Only as we enter the region of larger $\beta$, the curves start to deviate slightly while the qualitative behavior of the free energy is still the same for both values of $\kmax$.

Looking at the specific heat, in Fig.~\ref{fig:TRG_no_theta}, we see that truncation effects due to finite $\kmax$ are more pronounced. This is not too surprising, as the specific heat is obtained via Eq.~\eqref{eq:specific_heat} by approximating the second derivative with finite differences, 
\begin{align}
  \begin{split}
		&C(\beta,\theta=0) \approx \frac{\beta^2}{V}\frac{1}{12(\Delta \beta)^2}\times\\
    &\{16[\log\mathcal{Z}_{0}(\beta+\Delta\beta) - 2\log\mathcal{Z}_{0}(\beta)
		+ \log\mathcal{Z}_{0}(\beta-\Delta\beta)]\\
		&-[\log\mathcal{Z}_{0}(\beta+2\Delta\beta) - 2\log\mathcal{Z}_{0}(\beta)
		+ \log\mathcal{Z}_{0}(\beta-2\Delta\beta)]\},
	\end{split}
\end{align}
where we choose $\Delta \beta = 0.5$ for the data shown in the inset of Fig.~\ref{fig:TRG_no_theta}. Thus, the systematic errors in $\log\mathcal{Z}_{\theta}$ are enhanced by a factor of $(\Delta \beta)^2$ leading to significantly larger deviations in the regime of larger $\beta$. Focusing on smaller values of the inverse coupling, our data show that truncation effects are nevertheless small despite this effect, and we can reliably determine the specific heat with $\kmax=2$ in regimes $\beta\leq 1.1$. Hence, for all the following we will solely focus on that region and use $\kmax\geq 2$, for which systematic errors due to the truncation of the expansion in Eq.~\eqref{eq:expS} are very small.

We note that previous work using TRG~\cite{Kawauchi:2017dnj} observed large fluctuations in the specific heat in the regime of large $\beta$. In contrast, our ATRG results are stable even for large values of $\beta$. While Ref.~\cite{Kawauchi:2017dnj} used a bond dimension $D=21$, our calculations rely on a larger bond dimension $D=80$. Thus, we suspect that these fluctuations were caused by truncation effects due to the small bond dimension. 

\subsection{CP(1) model with $\theta$-term}
In order to study the CP(1) model in the presence of a $\theta$-term, we use the bond-weighted TRG method~\cite{2020arXiv201101679A} in conjunction with regular SVD to truncate the tensor at each coarse-graining step. Compared to the original TRG method, bond-weighted TRG introduces bond weights on the edges of the tensor network, allowing for more precise results within the same calculation time.

\begin{figure*}[ht!]
  \begin{center}
   \includegraphics[width=\textwidth]{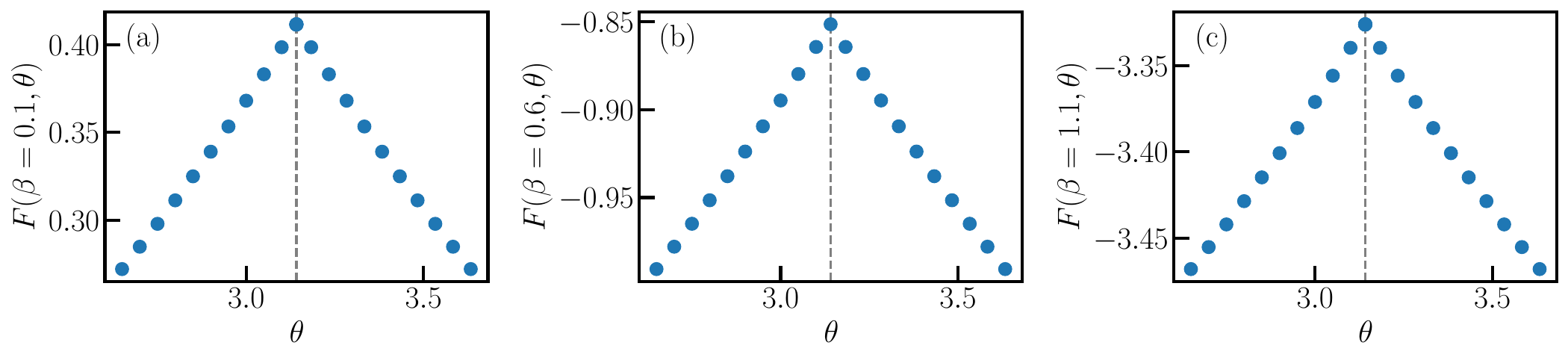}
  \end{center}
 \caption{TRG results for the free energy as a function of $\theta$, for the volume $V=2^{24}$, the bond dimension $D=80$, and the inverse couplings (a) $\beta = 0.1$, (b) $\beta=0.6$, and (c) $\beta = 1.1$. The vertical dashed gray line indicates $\theta=\pi$, at which we expect the phase transition to occur.} 
\label{fig:small_beta}
\end{figure*}

Figure \ref{fig:small_beta} shows our results for the free energy as a function of $\theta$ for the volume $V=2^{24}$, bond dimension $D=80$, $\kmax = 2$, a truncation of the character expansion of the topological term $\chi_\theta = 2$, and three values of the inverse coupling $\beta=0.1,0.6,1.1$. Focusing first on $\beta=0.1$ in Fig.~\ref{fig:small_beta}(a), we observe a clear cusp at $\theta=\pi$. Hence, the first derivative of $F(\beta=0.1,\theta)$ with respect to $\theta$ is discontinuous at $\theta=\pi$, indicating that the model undergoes a first-order phase transition at this point. This observation is consistent with previous findings of a first-order transition for $\beta = 0.1$~\cite{Plefka:1996ks,Azcoiti:2007cg,Kawauchi:2016dcg,Kawauchi:2017dnj}.

For larger inverse couplings of $\beta=0.6$ and $\beta=1.1$, the absolute values of the free energy change, as Fig.~\ref{fig:small_beta}(b) and Fig.~\ref{fig:small_beta}(c) reveal, respectively. However, for both $\beta=0.6$ and $\beta=1.1$, we still find the same cusp structure as for the smaller inverse coupling of $\beta=0.1$. Thus, even for these large values of $\beta$, the transition is still of first order. These results disagree with previous findings~\cite{Plefka:1996ks,Azcoiti:2007cg,Kawauchi:2016dcg,Kawauchi:2017dnj} that for $\beta_c \gtrsim 0.4-0.56$, the transition at $\theta=\pi$ becomes of second order and the first-order transition line bifurcates at $\theta\neq\pi$. 

\begin{figure*}[ht!]
  \begin{center}
   \includegraphics[width=\textwidth]{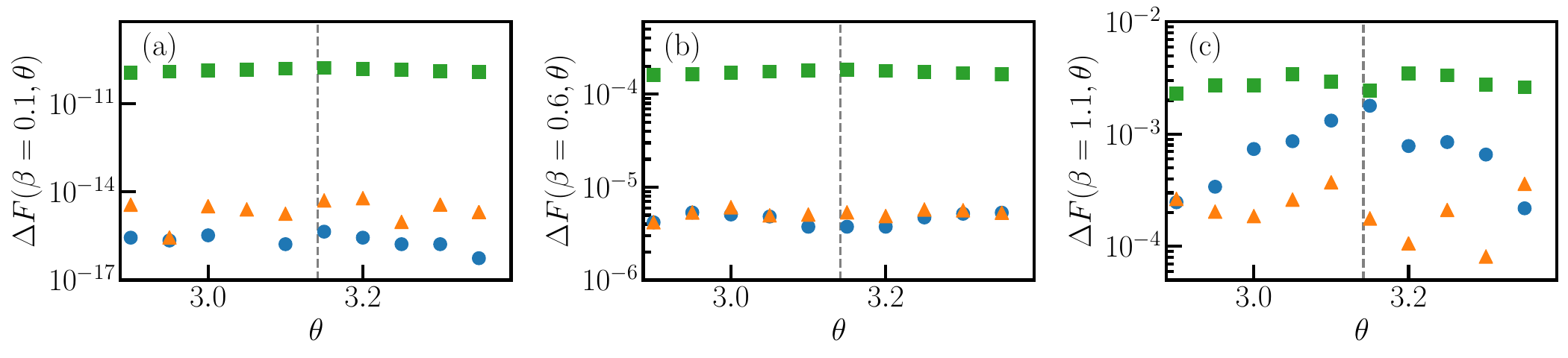}
  \end{center}
  \caption{TRG results for the systematic errors of computing the free energy as a function of $\theta$, for the inverse couplings of (a) $\beta=0.1$, (b) $\beta=0.6$, and (c) $\beta=1.1$. The blue dots in panels (a) and (b) show $\Delta F(\beta,\theta)$ for $D=112$, $\kmax=2$ and $\chi_\theta=2$, the orange triangles for $D=144$, $\kmax=3$ and $\chi_\theta=2$, and the green squares for $D=144$, $\kmax=2$ and $\chi_\theta=4$. For panel (c), the markers encode the same values of $\kmax$ and $\chi_\theta$, but we keep a fixed value of $D=112$ in all cases.}
	\label{fig:sys_err}
\end{figure*}

In order to ensure that the discrepancies for the phase structure obtained from our numerical results and previous studies are not caused by systematic errors due to the choice of $\kmax$, $D$, and $\chi_\theta$, we carefully examine the effect of these parameters. Figure~\ref{fig:sys_err} shows the difference in the results for the free energy, 
\begin{align}
\Delta F(\beta,\theta) = \left|F(\beta,\theta;D,\kmax,\chi_\theta) - F(\beta,\theta;80,2,2) \right|, 
\end{align}
between $(D,\kmax,\chi_\theta)=(80,2,2)$ and various larger choices of $\kmax$, $D$, and $\chi_\theta$ for a fixed volume of $V=2^{24}$. Note that changing $\chi_\theta$ and $\kmax$ affects the size of the initial tensor as discussed above. While for $(\kmax,\chi_\theta) = (2,2)$ the bond size of the initial tensor is $34$, for $(2,4)$ and $(3,2)$ it grows to $68$ and $98$ respectively. Since $D$ should be chosen well above the initial tensor size, we do not only increase $\kmax$ or $\chi_\theta$, but simultaneously also the value of $D$.

Figure~\ref{fig:sys_err}(a) shows our results for the smallest value of the inverse coupling corresponding to $\beta=0.1$. In particular, we observe that increasing the bond dimension while keeping $(\kmax,\chi_\theta) = (2,2)$ has an extremely small effect on the results, with an absolute change in the range below $10^{-14}$. Similarly, increasing $\kmax$ does not lead to noticeable deviations and the absolute change is in the same range. Increasing the value of $\chi_\theta$ to 4 yields a slightly larger deviation, which is still very small and in the range of $10^{-10}$. All in all, when we compare these absolute differences to the values of the free energy, which are on the order of $10^{-1}$ [see Fig.~\ref{fig:small_beta} (a)], the systematic errors due to finite $D$, $\kmax$ and $\chi_\theta$ are negligible throughout the entire range of $\theta$ we study. 

For the larger inverse couplings of $\beta=0.6$ and $\beta=1.1$ in Fig.~\ref{fig:sys_err}(b) and Fig.~\ref{fig:sys_err}(c), respectively, we observe a qualitatively similar picture. While the absolute deviation is larger than for the case of $\beta=0.1$, these effects are well below the percentile range even for our largest value of $\beta$. Just as before, increasing $D$ and $\kmax$ has the least impact, whereas increasing $\chi_\theta$ to $4$ now clearly has a slightly larger effect on the results. In relation to the values of the free energy shown in Fig.~\ref{fig:small_beta}(b) and Fig.~\ref{fig:small_beta}(c), which are on the order of $10^{-1}$ and 3, respectively, the systematic effects due to finite $(D,\kmax, \chi_\theta)$ are again negligible. 

\begin{figure*}[ht!]
  \begin{center}
   \includegraphics[width=\textwidth]{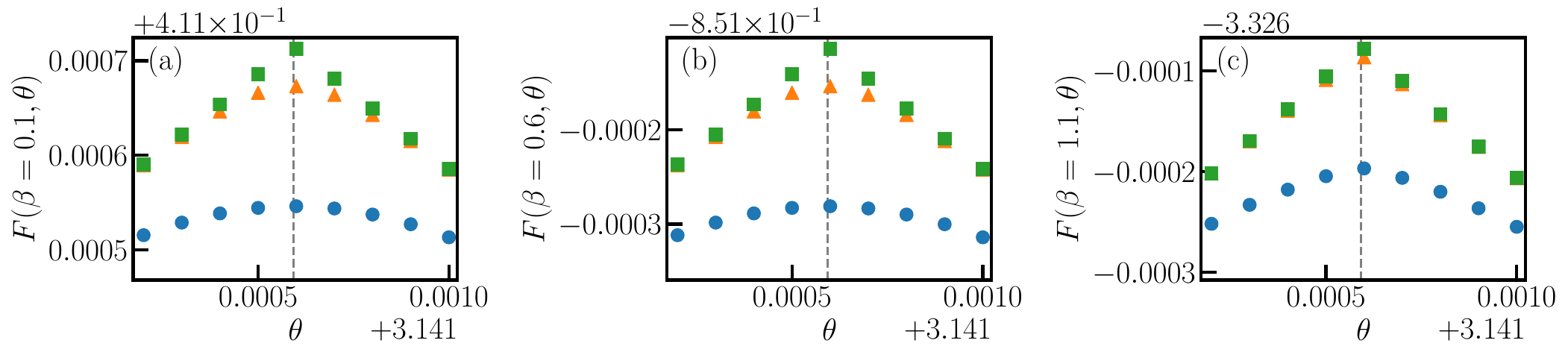}
  \end{center}
  \caption{TRG results for the free energy as a function of $\theta$ in a region very close to $\theta=\pi$ (indicated by the vertical gray dashed line), for the bond dimension $D=80$ and the inverse couplings (a) $\beta = 0.1$, (b) $\beta=0.6$, and (c) $\beta = 1.1$. The different markers correspond to the volumes $V=2^{12}$ (blue dots), $2^{14}$ (orange triangles), and $2^{24}$ (green squares).}
	\label{fig:free_energy_volume_dependence}
\end{figure*}

To further study the possible origin of the discrepancies between our and previous numerical results, we now investigate the cusp in the free energy as a function of the volume $V$. To this end, we plot a close-up around the region of $\theta=\pi$ for various volumes and various values of the inverse coupling in Fig.~\ref{fig:free_energy_volume_dependence}. Looking at our results for $\beta=0.1$ [see Fig.~\ref{fig:free_energy_volume_dependence}(a)], we observe that for volumes below $2^{14}$, the free energy is smooth at $\theta=\pi$. Thus, the first derivative does not show any discontinuity for small volumes. Only when increasing $V$ to larger values, the cusp at $\theta=\pi$ eventually emerges, signaling the onset of the first-order transition directly in the free energy. For larger values of the inverse coupling, corresponding to $\beta=0.6$ and $1.1$, we see a qualitatively similar behavior, as Figs.~\ref{fig:free_energy_volume_dependence}(b) and \ref{fig:free_energy_volume_dependence}(c) reveal. Interestingly, the finite-volume effects seem to be slightly less pronounced for larger values of $\beta$, as a comparison between Figs.~\ref{fig:free_energy_volume_dependence}(a) and \ref{fig:free_energy_volume_dependence}(c) shows. For $\beta=1.1$, we observe that the data for $V=2^{14}$ already has a clear signature of a cusp, and increasing the volume only leads to a slightly sharper peak.

In summary, the analysis of the systematic errors from finite $(V, D,\kmax, \chi_\theta)$ further supports the reliability of our findings throughout the entire range of $\beta$ that we study. We conclude that the phase transition at $\theta = \pi$ is still of first order up to an inverse coupling of $\beta = 1.1$.

\section{Discussion and Conclusions\label{sec:conclusion}}

In this paper, we numerically studied the phase diagram of the lattice CP(1) model with a $\theta$-term. We observe a CP-violating, first-order phase transition at $\theta=\pi$ and small $\beta$, in agreement with previous findings~\cite{Plefka:1996ks,Azcoiti:2007cg,Kawauchi:2016dcg,Kawauchi:2017dnj}. However, our results do not confirm previous indications for a bifurcation in the first-order transition line at some critical value of $\beta$, ranging from $\beta_c=0.56$~\cite{Plefka:1996ks} to $\beta_c = 0.5$~\cite{Azcoiti:2007cg} to $\beta_c=0.4$~\cite{Kawauchi:2016dcg,Kawauchi:2017dnj}. Instead, we find that the first-order transition at $\theta=\pi$ persists up to $\beta=1.1$ without any bifurcation, and we observe no indication that the phase transition at $\theta=\pi$ becomes of second order for $\beta>\beta_c$. The existence of a second-order transition at $\theta=\pi$ and $\beta>\beta_c$ was suggested by the previous studies based on a strong coupling analysis~\cite{Plefka:1996ks}, Monte Carlo simulations~\cite{Azcoiti:2007cg}, and first TRG studies~\cite{Kawauchi:2016dcg,Kawauchi:2017dnj}. 

Compared to the strong coupling analysis of the CP(1) model in Ref.~\cite{Plefka:1996ks}, our method does not require an expansion in the inverse coupling $\beta=1/g^2$, but instead performs truncations and character expansions to construct the tensor representation. Thus, while for small $\beta$ the results in~\cite{Plefka:1996ks} may be more accurate than ours, the strong coupling analysis becomes more challenging for larger values of $\beta$. Since our TRG results clearly show that the effects due to the truncations of the character expansions are very small throughout the entire range of $\beta$ we study, our results suggest that the strong coupling expansion does not correctly capture the physics of the model up to $\beta=1.1$.

Similarly, the Monte Carlo studies in Ref.~\cite{Azcoiti:2007cg} suffer from the sign problem, which makes it difficult to reliably examine the phase diagram of the CP(1) model for large values of $\theta$. In particular, the Monte Carlo approach becomes more challenging when approaching the phase transition at $\theta=\pi$, whereas our TRG calculation only shows small systematic errors, which are essentially independent of the value of $\theta$ for the parameter range we study.

Compared to previous TRG-based studies of the model, our approach to determine the phase structure is more direct. In particular, Refs.~\cite{Kawauchi:2016dcg,Kawauchi:2017dnj} did not study the free energy of the model at large volumes, but rather explored the free energy at small volumes and used finite-size scaling behavior of the topological susceptibility as an indicator to determine the order of the phase transition. In contrast, our computations are performed at a much larger volume, for which the results are well converged, which allows us to study the behavior of the free energy without having to rely on finite-size scaling. Indeed, as we have shown in Fig.~\ref{fig:free_energy_volume_dependence}, a clear signature of the first-order transition only occurs at volumes of $V>2^{14}$. We note that our bond dimension of $D=80$ is only slightly larger than the bond dimension $D=68$ used in Ref.~\cite{Kawauchi:2017dnj}, but increasing the bond dimension only has a minor effect on our results, see Fig.~\ref{fig:sys_err}.

To summarize, our results demonstrate that the phase transition at $\theta=\pi$ is of first order up to $\beta\leq 1.1$, and no second-order transition occurs. The discovery of such a second-order transition would be crucial for confirming Haldane's conjecture~\cite{Haldane:1982rj,Haldane:1983ru} that the O(3) model becomes gapless at $\theta=\pi$ and weak coupling. As our study reveals, the possibility of a second-order transition at $\theta=\pi$ and a bifurcation of the first-order transition line at $\theta\neq\pi$ is only given for $\beta_c>1.1$. Compared to previous studies, our method does not suffer from the sign problem as Monte Carlo simulations, and our results have small systematic errors of $\lesssim 10^{-3}$ due to larger volumes than in earlier TRG studies. Thus, our study reveals that testing Haldane's conjecture in the weak-coupling regime requires both major numerical efforts and detailed systematic error analysis. In the future, we will extend our study to larger values of $\beta>1.1$. While this task is more challenging due to enhanced truncation effects in the weak-coupling regime, our numerical make us positive that this parameter range can be reliably accessed with TRG methods.

\begin{acknowledgments}
  Research at Perimeter Institute is supported in part by the Government of Canada through the Department of Innovation, Science and Industry Canada and by the Province of Ontario through the Ministry of Colleges and Universities.
  Y.J.K.\ acknowledges financial support from the Ministry of Science and Technology (MOST) of Taiwan under grants No.\ 110-2112-M-002-034-MY3 and 108-2112-M-002-020-MY3.
  S.K.\ acknowledges financial support from the Cyprus Research and Innovation Foundation under project ``Future-proofing Scientific Applications for the Supercomputers of Tomorrow(FAST)'', contract No.\ COMPLEMENTARY/0916/0048.
\end{acknowledgments}

\appendix  

\section{Finite-size scaling for the susceptibility}
\begin{figure}[H]
  \includegraphics[width=0.95\columnwidth]{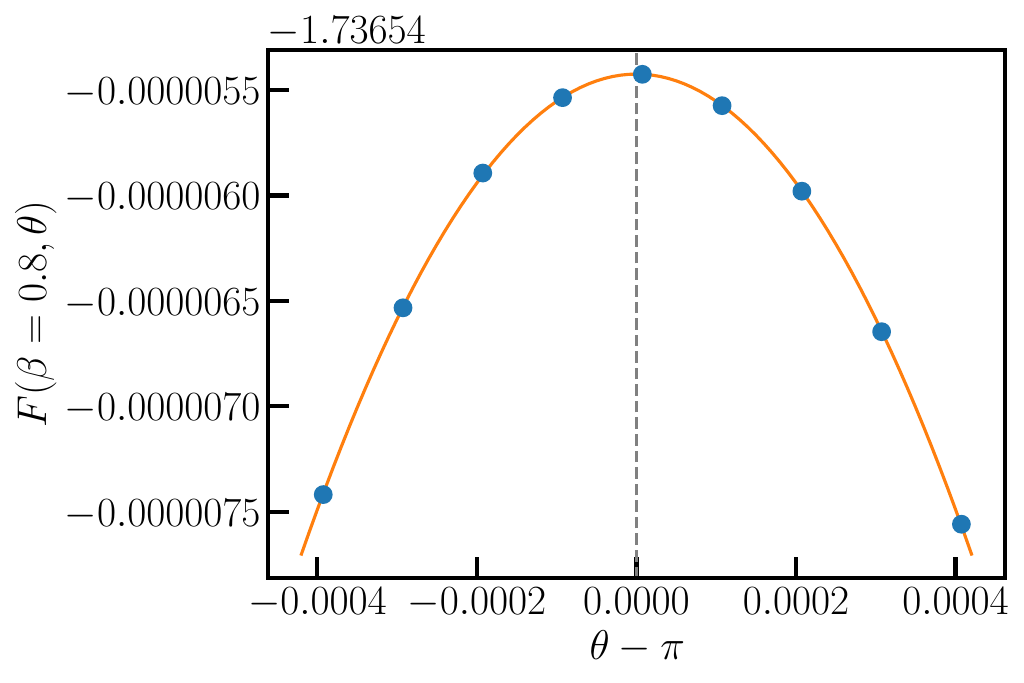}
  \caption{TRG data for the free energy for $\beta=0.8$, $V=2^8$ (blue dots) and fit to our numerical data according to Eq.~\eqref{eq:polynomial_interpolation} (orange solid line) around $\theta = \pi$ (vertical gray dashed line).}
  \label{fig:susceptibility_interpolation}
\end{figure}

In the main text, we focused on the shape of the free energy and used the emergence of a cusp toward the thermodynamic limit as an indication for a first-order phase transition. Alternatively, following Refs.~\cite{Kawauchi2016a,Kawauchi2018}, we can also examine the finite-size scaling of the peak value $\chi_\mathrm{peak}$ of the susceptibility
\begin{align}
  \chi = \frac{1}{V}\frac{\partial^2 \log \mathcal{Z}_\theta}{\partial \theta^2} = -\beta \frac{\partial^2 F(\beta,\theta)}{\partial \theta^2}
  \label{eq:susceptibility}
\end{align}
at $\theta=\pi$. For a first-order phase transition, it is expected that $\chi_\mathrm{peak}$ is proportional to the volume, whereas for a second-order transition, one expects a scaling of $\chi_\mathrm{peak}\propto V^\gamma$ with $\gamma<1$~\cite{Privman1990}. In principle, the susceptibility could be directly determined from our TRG results for $\mathcal{Z}_\theta$, by computing the derivative numerically. However, this would require a very fine resolution in order to avoid large errors. Thus, we follow Ref.~\cite{Kawauchi2018} and approximate our data for the free energy close to $\theta=\pi$ with a polynomial. Since finite-size effects round off the peak in the free energy, we choose the functional form
\begin{align}
  F(\beta,\theta) \approx c_0 + c_1(\theta-\pi)^2 + \mathcal{O}\left((\theta-\pi)^4\right),
  \label{eq:polynomial_interpolation}
\end{align}
where we consider only even powers to ensure the symmetry around $\theta = \pi$. An example of this interpolation is shown in Fig.~\ref{fig:susceptibility_interpolation}.
The peak value of the susceptibility can then be obtained from the polynomial interpolation using Eq.~\eqref{eq:susceptibility}, and is given by $\chi_\mathrm{peak} = -2\beta c_1$.

\begin{figure*}
  \includegraphics[width=1.0\textwidth]{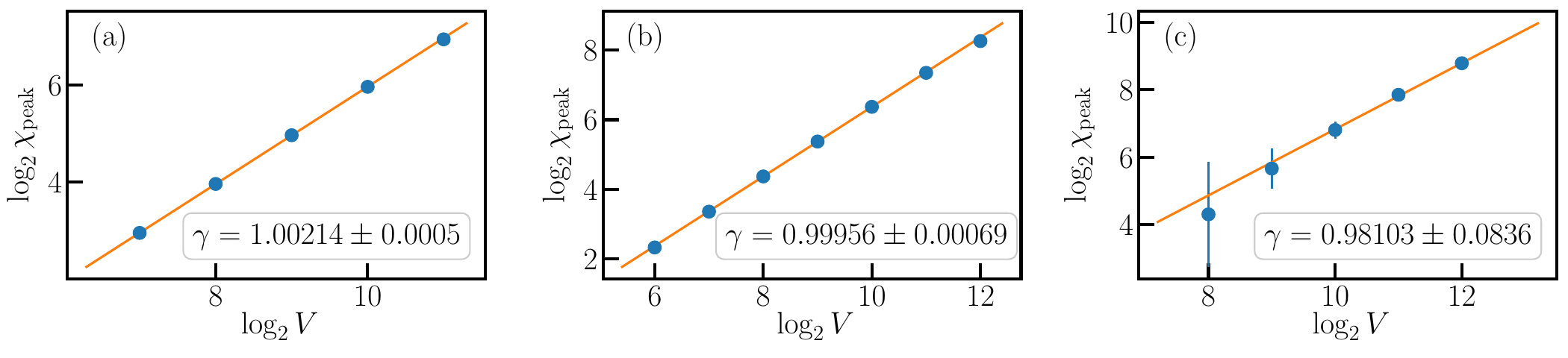}
    \caption{Logarithm of the peak value of the susceptibility (blue dots) as a function of the logarithm of the volume for (a)~$\beta=0.6$, (b) $\beta=0.8$ and (c) $\beta=1.1$. The error bars on the data points represent a systematic uncertainty resulting from determining the peak value of the susceptibility. The orange solid lines correspond to linear fits to the data points, and the boxes in the right lower corner show the resulting slope of the fit.}
    \label{fig:susceptibility_vs_volume}
\end{figure*}

In order to determine the volume dependence of $\chi_\mathrm{peak}$, we repeat the analysis described above for various volumes. The values for the volume are chosen such that they are large enough to avoid excessive finite-size effects
\noindent and small enough to avoid a cusp in the free energy, as we observed for large volumes in the main text. Our results for this analysis for various values of the coupling are shown in Fig.~\ref{fig:susceptibility_vs_volume}.
Focusing on $\beta=0.6$ first [see Fig.~\ref{fig:susceptibility_vs_volume}(a)], we clearly observe that $\chi_\mathrm{peak}$ is proportional to the volume and $\gamma$ is in good agreement with $1$.
Increasing the inverse coupling to $\beta = 0.8$, we see a very similar picture, as the data in Fig.~\ref{fig:susceptibility_vs_volume}(b) reveals. Again, the peak value of the susceptibility scales linearly with the volume, and the value for $\gamma$ obtained from our fit is compatible with $1$ within error bars.
For $\beta=1.1$, we observe that it becomes increasingly harder to find a window where finite-size effects are not excessive and the free energy is still far enough away from showing a cusp. As a result, the data points for small values of the volume in Fig.~\ref{fig:susceptibility_vs_volume}(c) show slightly larger error bars than for the previous values of $\beta$. When determining the volume dependence of $\chi_\mathrm{peak}$ for this case, we again obtain a value of $\gamma$ that agrees with $1$ within error bars.

In summary, the finite-size scaling of the peak value $\chi_\mathrm{peak}$ of the susceptibility indicates that the phase transition is of first order throughout the entire range of $\beta$ under consideration. This is consistent with our results in the main text, as obtained from the direct examination of the free energy in the limit of large volumes.

\bibliography{Papers}
\end{document}